# Multiple Model Software for Airflow and Thermal Building Simulation. A case study under tropical humid climate, in Réunion Island.

H. BOYER        *Université de la Réunion(\*) / INSA de Lyon (\*\*)*
J. BRAU         *CETHIL, INSA de Lyon (\*\*)*
J.C. GATINA     *Université de la Réunion (\*)*

*Abstract :* The first purpose of our work has been to allow -as far as heat transfer modes, airflow calculation and meteorological data reconstitution are concerned- the integration of diverse interchangeable physical models in a single software tool for professional use, *CODYRUN*. The designer's objectives, precision requested and calculation time consideration, lead us to design a structure accepting selective use of models, taking into account multizone description and airflow patterns. With a building case study in Reunion Island, we first analyse the sensibility of the thermal model to diffuse radiation reconstitution on tilted surfaces. Then, a realistic balance between precision required and calculation time leads us to select detailed models for the zone of main interest, but to choose simplified models for the other zones.

I ) Display of the simulation tool :

Born from a joint research project involving both the Université de la Réunion and INSA de Lyon, this work aims at producing an efficient building thermal simulation tool, including some research and conception aspects, and taking into consideration different types of climates. More precisely, it is a multizone software integrating both natural ventilation and moisture transfers, called *CODYRUN*.

a) Architecture and original aspects

The three main parts are the building description, the simulation program and the operation of the results. As far as the description is concerned, we have been brought up to break down the building into three types of entities which are the following ones : firstly, the *Zones* (from a

(\*) *Laboratoire de Génie Industriel, Université de la Réunion - Faculté des Sciences, 15 rue Cassin 97489 Saint-Denis Cedex. Ile de la Réunion-FRANCE*

(\*\*) *Institut National des Sciences Appliquées. Equipe Equipement de l'Habitat, CETHIL, Bat. 307, 20 Avenue Albert Einstein. 69621 Villeurbanne Cedex, FRANCE* thermal point of view), the *Inter-zones* partitions (zones separations, outdoor being considered as a particular zone) and finally, the *Components* (i.e. the walls, the glass partitions, air conditioning system sets, so on and so forth). For a simulation, the *Project* notion includes a weather input file, a building and then, a result file. The following treelike structure illustrates our organisation :

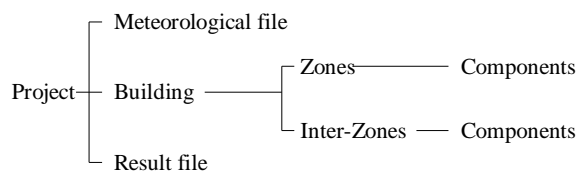

Fig. 1 *: Data organization*

During a simulation, one of the most interesting aspects is to offer the expert thermician a wide range of choices between different heat transfer models and meteorological reconstitution parameter models. The aim of this simulation may be, for a given climate, to realize studies of the software sensitivity to these different models, in order to choose those that should be integrated in a suitable conception tool.

In the second part of this article, the first application concerns the comparison of two models of sky diffuse radiation. In the same way, for a given climate, depending on the objective sought during the simulation (through tracking the temperatures or estimation of a yearly energy consumption), it is interesting to have the liberty of selecting the models to be involved.

In most existing simulation software, the choice of the models being already done, their application is global to the whole building. If these models are complex, the multi-zone feature as well as airflow patterns leads quickly to calculation time which is not compatible with a conception tool. It has seemed interesting to us to allow, for some of the phenomena, a selective use of the models. Thus, the choice of the indoor convection model is realized during the definition of a zone or else, the choice of the conduction model is done during the description of a wall. The aim is then to link the level of complexity of the models concerning one entity to the interest beared for this same entity. And that is how we will show in the second part of this article the importance in choosing a detailed model for the indoor convection in the main interest zone and simplified models for the other zones.

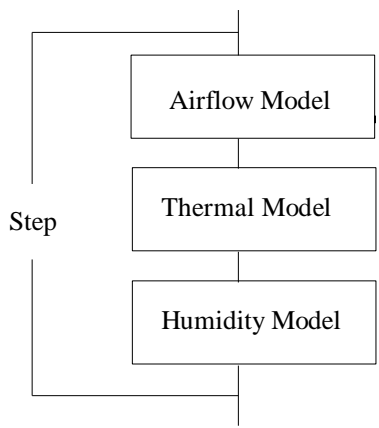

Fig. 2 : *General flowchart*

In relation with calculation program, the software executes at each step the reckowing of airflow patterns, temperature field and specific humidity of each zone.

The most simplified airflow model considers as known the airflow rates between all zones, whereas the more detailed model calculates with a pressure model the airflow through each of the openings, which can be large ones. The building is represented as a network of pressure nodes, connected by non-linear equations giving the flows as a function of the pressure difference. This detailed airflow calculation goes through the iterative solution of the system of non linear equations made up with air mass conservation inside each zone.

The thermal model relies on INSA's previous simulation code, *CODYBA* (BRAU, 1987). With the usual physical assumptions, we use the technique of nodal discretisation of the space variable by finite difference. In addition, the mass of air inside one zone is represented by a single thermal capacity. Thus, for a given zone, the principle of energy conservation applied to each concerned wall node, associated with the sensible balance of the air volume, constitute a set of equations, that can be condensed in a matricial form :

$$ C \frac{dT}{dt} = A \cdot T + B \qquad (1)$$

The automation of the setting up of the previous equation requires a decomposition of A into twelve elementary matrices, each one having a particular feature.

$$ C \frac{dT}{dt} = $$
$$ [A_{cond} + A_{cvi\_lin} + \ldots + A_{connex}] \cdot T + $$
$$ B_{int\_load} + \ldots + B_{connex} \qquad (2)$$

Thus, $A_{cond}$ gathers features linked to heat conduction through the walls whereas $A_{cvi\_lin}$ gathers the features depending on indoor and linearised convective exchanges. In the same way, for each step, the filling up of B vector is made easier by its decomposition into fifteen elementary vectors. The physical coupling of the considered zone to the other ones is realized through an iterative connection process, through the filling up of a matrice $A_{connex}$ and a vector $B_{connex}$.

At each step, the resolution of equation (1) uses an implicit finite difference procedure and the coupling iterations between the different zones make it possible to calculate the evolution of temperatures as well as those of sensible powers needed in case of air conditioning. Having in mind a compromise precision/calculation time, it is to be noticed that the thermician using the software keeps the control of the solving methods, of the iterations (mainly concerning connection process and airflow model) and of the different convergence criterions.

### b) WINDOWS Front-End :

Developed on PC micro computer with Microsoft WINDOWS interface, implemented in C-language, the software beneficts of all the user-friendliness required for a conception tool (windowing, mouse, ...).

A more technical aspect linked to this system is very interesting for simulation tools based on PC, such as our software : memory managing supplied by WINDOWS allows to allocate much more than the classical 640 Ko limit, which is most necessary as regards the capacity and the variety of the matrices and vectors that interfere in the simulation process.

The user's interface proposes to begin the description of a building with the following window :

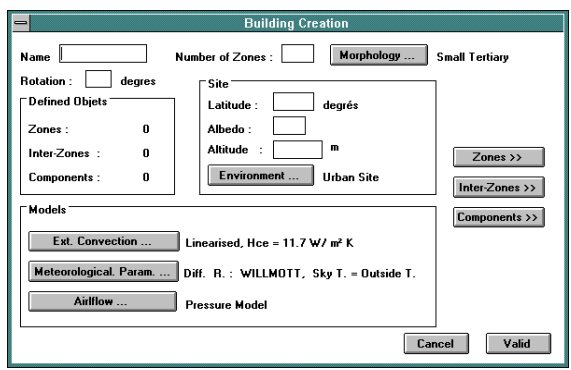

Fig. 3 : *Building description window*

The push buttons zone, interzone and components make possible the access to this previously defined entities. The access to the models linked to the entity "building" is also possible from this window through the push buttons of the screen part called "models". It is possible in this way to select the chosen models for outside heat convection transfer, reconstitution of meteorological parameters (diffuse and sky temperature) as well as the airflow model.

## II ) Case study :

### a) Description :

Réunion Island is located in the Indian Ocean by 55°1 longitude East and 21°5 latitude South. The climatic conditions being those of a tropical and humid climate, we have reconstituted an hourly file gathering all the meteorological parameters necessary for our simulations, i.e. mainly solar radiation data, outdoor dry temperature and moisture rate, and wind. With *CODYRUN*, for some periods of the considered year, we propose to reproduce a building indoor thermal conditions.

The chosen building is a cubic shape with a side of 6 m, in contact with the ground. We consider a thermal split up in three zones : the ground floor, the eastern first floor and the western first floor. The following sketch illustrates our description :

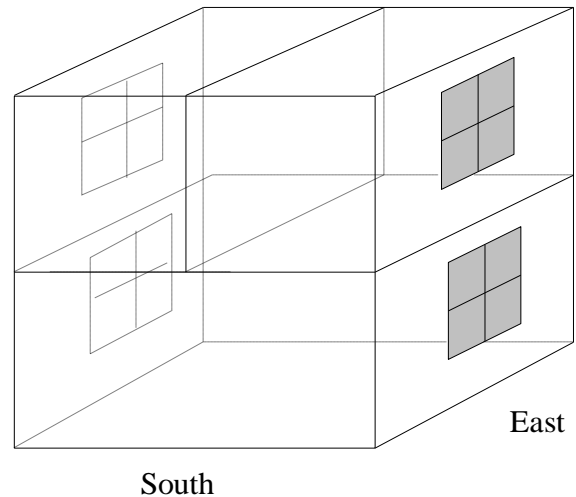

Fig 4 : *Building sketch, Southern face*

The main characteristics are the following ones : let us suppose all the walls are made up with of dense concrete of 12 cm, the slab on grade being of 30 cm in the same material. On the eastern and western sides are display bay windows (simple glass) of an elementary surface of four square meters and of usual optical and conductive quality**.** Later, the conductive model kept for each wall and glass is a simple two capacitors model (R2C).

Besides, we suppose the presence of an air conditioning system in the western zone, upstairs, but we shall come back on this component in a following paragraph. The building is described with three zones, sixteen inter-zones and twenty two components.

### b) Diffuse radiation model choice :

Under tropical-humid climate, the readings made on the short wave diffuse radiation show the importance of this kind of inputs. A quick analysis of the graphs on the site of Réunion Island shows an important diffuse radiation when the direct beam is low**.** Tropical humid climate leads to take into account the total diffuse radiation (sky diffuse and ground reflection), when designing the protection devices, i.e. screens, shadow masks, ... (Cabirol, 1984).

Most of solar energy calculations consider this diffuse radiation as being isotropic. As given *dh* the diffuse and horizontal radiation, on a tilted plane (azimuth $\gamma$, inclination *s*), this diffuse radiation is :

$$d(s, \gamma) = (\frac{1+\cos s}{2}).dh \quad (W/m^2) \quad (3)$$

Meanwhile, anisotropic models having been validated (Gueymard, 1987), we have integrated the Willmott model (Lebru 1983), besides the isotropic model for reconstitution of incident diffuse radiation on a tilted plane. With the following notations, the proposed expression is :

$$d(s,\gamma) = \quad (4)$$
$$(F.C(s).\frac{1+\cos s}{2} + (1-F).\frac{Max(\cos i, 0)}{\sin h}).dh$$

with $C(s) = 1.00115 - 3.54.10^{-2} s - 2.46.10^{-6} s$

$$F = 1 - \frac{Gh}{GhExt}(1 - \frac{dh}{Gh})$$

and

- *i*      mean angle of incidence of beam radiation with respect to the surface normal
- *h*      solar altitude
- *Gh*      site horizontal global radiation
- *GhExt* extraterrestrial global radiation

Considering the sharing (between direct beam and diffuse part) of the solar heat gains entering a zone (through the windows), the model considers that the direct beam is incident on the floor and that the diffuse part is split up depending on the proportion of the surfaces. Through the optical characteristics of the materials in which the surfaces of the walls in consideration are made up, it is necessary to solve a linear set of equations to obtain the radiation absorbed by the surface nodes of the walls. These radiations are constituting one of the elementary vector, B*swi*.

For the study, we have taken from the yearly file a sequence of two extreme consecutive days, one cloudy and the other sunny. The following graphs show the evolution of the horizontal direct beam and diffuse radiation for the simulation period.

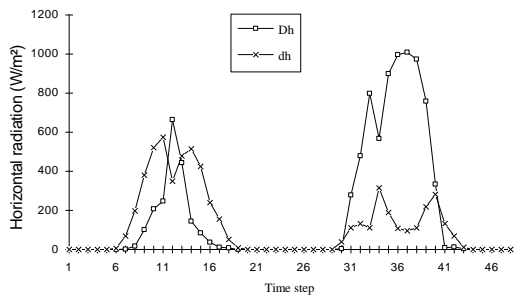

Fig. 5 : *Horizontal radiation*

Through the glasses, one part of the energy load is due to the beam part of the transmitted radiation and the other part to its diffuse part. For the first day, the following graphic plots show the evolution of the total diffuse incident radiation in the eastern floor zone, cases isotropic and anisotropic (Willmott).

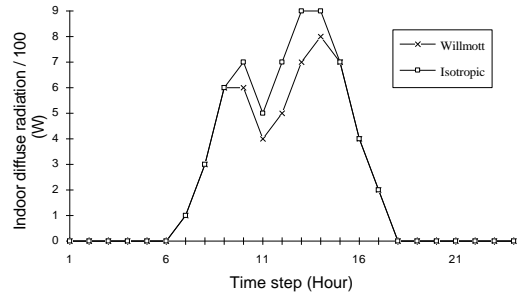

Fig. 6: *Eastern floor diffuse radiation*

To integrate a non isotropic diffuse radition pattern is the same as to consider as non isotropic a part of the diffuse radiation preliminary calculated with the isotropic model. As a consequence, the diffuse radiation incident in a zone will be lesser with the anisotropic model, as shown by the previous graph. For the incident direct beam, the live would show the opposite order.

The following figure allows comparison between graphs of dry temperatures evolution concerning the eastern first floor, and that for the two cases previously mentioned.

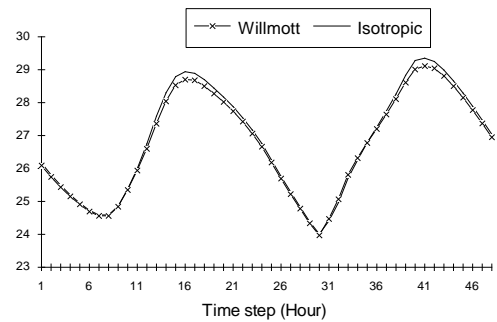

Fig. 7 : *Eastern floor dry temperature*

Considering our building and the mentioned period, the graphs are practically merge. This simulation shows in our particular case that the whole building model is little sensitive to the sky diffuse radiation model. A simple interpretation of this phenomenon can be delivered. In the anisotropic case, the diffuse radiation deriving from the isotropic model is divided into two parts, one of them is isotropic, the other one directing. As far as the energetic behaviour of a building is concerned, the coupling of the external short wave radiation and the zone in consideration is realized through the glasses. For simple type of glass such as the one in which our windows are made, the bibliographic expressions of the transmittance for direct or diffuse radiation are but little different. Moreover, the diffuse transmittance is not dependent of the incidence angle. Thus, the difference in dividing the energy load according to the isotropic criterion will have but little incidence on the zone's thermal behaviour.

### c ) Choice of the indoor convection model

#### c -1) HVAC System Component :

The western zone being air conditioned, for a better understanding of the following graphs, we first display and comment the data window capture of the *Air Conditioning System* component.

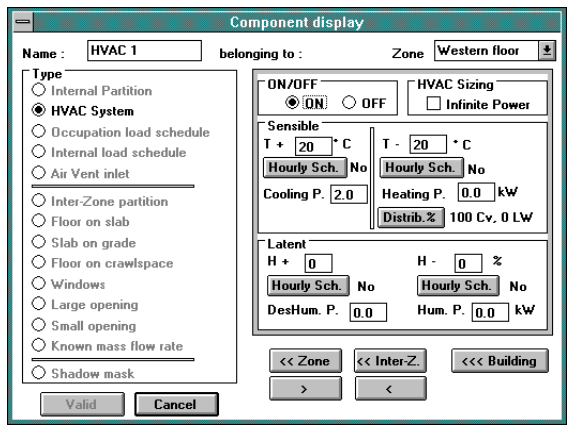

Fig 8 : *Component description window*

Thus, for an air conditioning system, information relative to the sensible and latent loads are to be entered. For our study, we'll consider the sensible part. We must then define the threshold temperature values (high and low), the hourly schedules and the available heating and cooling powers. Most of the time, simulation softwares consider the heating power as convective. However, to allow a better integration of systems such as heating floors, it is possible to choose for the involved power convective and radiative ratios. In particular case of the air conditioning system sizing, the software considers that the available power is infinite.

#### c -2) The indoor convection

The choice of the indoor convection model is made during the zone description. We have set up three models as regard the indoor convection.

The first one being a linear model with a coefficient of constant convective exchanges (but that can be modified by the conceptor), the second one, linear with a coefficient depending on the type of wall (i.e. floor, ceiling, vertical wall) and the last model is a non linear one. In this case, the exchange coefficient depends in a non linear way on the temperature difference between air and the given surface. The integration of such a model in our system goes then through the filling up of a vector dealing with non linear and convective flows, $B_{cvi\_lin}$ and through a process of iterative resolution with a convergence criterion that can be modulated for example as regards the indoor dry temperature. This non-linear model needs several resolutions of the equation (1), at each step. With a convergence criterion of $10^{-3}$ °C on the air temperature of the zone in consideration, the number of iterations is three or four. The use of this model tends then to penalize the tool as far as the calculation time is concerned, but this model can be considered as the most reliable.

#### c -3) Exploitation

Considering the conditions of simulation of the previous a) paragraph, our concern is now the evolution of the dry temperature inside the western zone floor as well as the sensible power needed. Firstly, we consider the A case in which the convection integrated with the help of a linear model with a constant coefficient (5 W/m².K), in each zone. The following graphs are obtained :

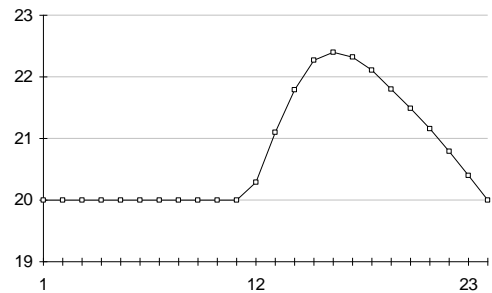

Fig. 9 : *Air Temperature, Western floor*

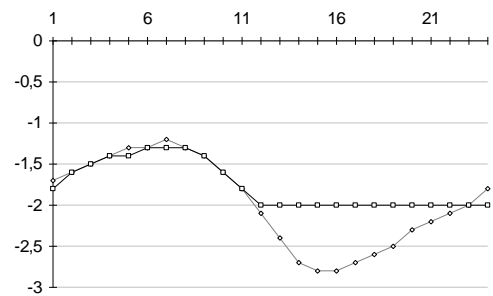

Fig. 10: *AC Power, Western floor*

The graph of the power in dotted corresponds to equipment sizing, in which case we consider the air conditioning power as infinite. This graph enables us to determine the cooling power to be set up, about 3kW, in order to respect at any time the specified 20°C.

If the cooling equipment power is inferior, (2 kW in our case), the result is an increase of the indoor temperature during the second half of the day (Fig. 9). Quantitatively, the two previous graphs allow a designer to realize the overheating (and its length) due to the under-sizing of the air conditioning system.

In case B, we integrate the indoor convection as non linear, for the three zones. This simulation is very costly as far as calculation time is concerned. In the last case, C, we'll deal with the non linear convection only in the zone in which we are most interested, i.e. the western floor, and we'll use the linear model with a constant coefficient for the two other zones.

The superposition of the graphs relative to the three cases produces the following results.

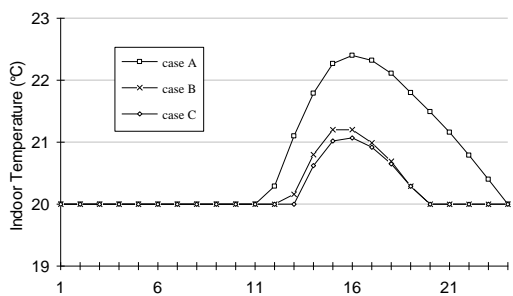

Fig. 11: *Air Temperature, Western floor*

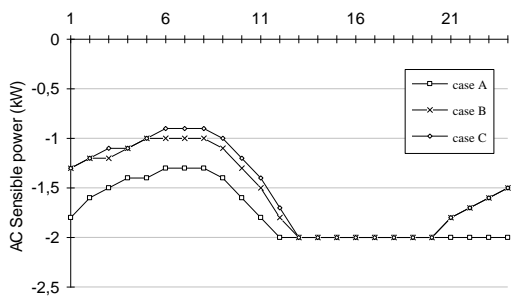

Fig. 12: *AC Power, Western floor*

If we consider as right the values proposed for case C, the inaccuracy of the model with a constant coefficient applied to all the zones is clearly visible. The mistake on the temperature reaches more than one degree and these on the power needed is of 0,5 kW. On the contrary, the case C shows the accuracy of the selective application of the non linear convection model,
with graphs very close to the case B, both in temperature and in power. In parallel, the comparison of the simulation times (outside building initialization period) for a given period of the day is given in the following board.

| Case | Temp. Err. (°C) | Power Err. (kW) | Time (mn"s) |
|------|-----------------|-----------------|-------------|
| B    | 0 (reference)   | 0 (reference)   | 2"53        |
| A    | 1.2             | 0.5             | 0"54        |
| C    | 0.1             | 0.1             | 1"35        |

Thus, there is approximativaly a one to a half ratio between the simulation time in case B and C. This ratio can be found again by other simple considerations. Indeed, it is possible in first approximation to suppose as constant the time necessary to set up and solve the state equation of one zone, i.e. $t$ seconds. If *Nb_zones* is the number of zones, in the case A, the resolution for one step requires $t_A = (Nb\_zones \cdot t)$ seconds. Let us suppose constant and equal to $i$ the number of iterations in a zone, interations introduced by the non linear model of the indoor convection. In case B, all the different zones being concerned with the non linearity, the calculation time equals $t_B = (Nb\_zones \cdot i \cdot t)$ seconds. In case C, this time becomes :

$$t_C = (i + (Nb\_zones - 1) \cdot t) \quad \text{seconds.}$$

With values corresponding to our case (*Nb_zones* = 3, i = 3), the ratio ($t_C/t_B$) is quite close to 0.5.

## III) Conclusion

Through the previous simulations, we have introduced *CODYRUN*, focusing on a few points. Many aspects developed to this day in this software have been purposely put aside, in particular in regard to airflow simulation, which is a dominant heat transfer mode under tropical climate. As far as heat conduction transfer is concerned, the different implemented models haven't been reviewed. In the same way, as regard this transfer mode, tools such as physical aggregation of walls or modal reduction (Roux 1988), allowing a notable decreasing of calculation time, will be the objects of future improvements. So, we join one of the preoccupations of his study, i.e. the balance between precision required and calculation time.

REFERENCES

Auffret, P.; Gatina, J.C.; Hervé ,P. 1984. "Habitat et climat à la Réunion. Construire en pays tropical

humide." Documents et recherches n°11. Université de la Réunion.

Brau, J.; Roux, J.J.; Depecker., P.; Croizer, J.; Gaignou, A.; Finas, R. 1987."Micro-informatique et comportement thermique des bâtiments en régime dynamique." *Génie Climatique*, 1987, n°11 (Oct-Nov), 15-23

Cabirol, T. ; 1984; "Habitat bioclimatique : l'incidence de l'ensoleillement et du vent". *Afrique Expansion*, 1984, n° 6 (Nov. ), 45-48

Lebru, A. 1983. "Estimation des irradiations solaires horaires dans un plan quelconque à partir de la donnée de l'irradiation horaire globale (et éventuellement diffuse) horizontale." Research Report N° 239 cahier 1847. CSTB, Sophia Antipolis FRANCE.

Roux, J.J.; Depecker; P., Krauss, G.; 1988. "Pertinence and performance of a thermal model adapted to CAD context." In *Proceedings of the sixth International PLEA Conference* (Porto, Portugal, July 27-31),749-754

Geymard, C.; 1987; "An anisotropic solar irradiance model for tilted surface and its comparison with selected engineering algorithms." *Solar Energy*, vol. 38, N° 5, 367- 386

Walton, G.N.; 1984; National Bureau of Standards, Washington, DC. "A computer algorithm for predicting Infiltration and Interroom Airflows". *ASHRAE Transactions* 84-11, N°3